\documentclass{optica-article}
\journal{opticajournal} 

\articletype{Research Article}
\usepackage{soul}
\usepackage{lineno}

\begin{document}

\title{Controlling the degree of entanglement in downconversion by targeted birth zone activation}

\author{Vikas S Bhat, Rounak Chatterjee, Kiran Bajar and Sushil Mujumdar\authormark{*}}

\address{Nano-Optics and Mesoscopic Optics Laboratory, Tata Institute of Fundamental Research, 1, Homi Bhabha Road, Mumbai, 400 005, India}

\email{\authormark{*}mujumdar@tifr.res.in} 


\begin{abstract*} 
We explore the consequences of varying the pump beam waist that illuminates a nonlinear crystal, realizing spontaneous parametric down-conversion (SPDC). The coherence is transferred from the marginal one-photon wavefunction to the two-photon wavefunction where it manifests into entanglement in the form of spatial correlation. 
We interpret this as a consequence of the number of independent emitters, called the biphoton birth zones, targeted by the pump beam on the crystal. The birth zone number $N$ characterises the number of such birth zones that fit along a diameter of the region illuminated by the pump waist. 
To experimentally observe the duality between  the one- and two-photon interference, we employ a double slit and analyse their visibilities $V_m$ and $V_\text{12}$ respectively. 
We demonstrate the conservation of the quantity $V_m^2+V_\text{12}^2$.
Finally, we identify three regimes of entanglement of the down-converted photons based on $N$. We show that changing the pump waist lets us actively control the degree of entanglement letting us access these regimes. We provide implications of each regime, and mention experimental use cases thereof. 

\end{abstract*}

\section{Introduction}\label{sec:introduction}
    The field of two-photon interferometry was born out of some pioneering experiments done in the late 80s \cite{hom,rupaGhosh, shih} and early 90s \cite{zwm, Ou1990, rarity} when entangled photons generated out of SPDC were used. They resulted in a theoretical exposition by Horne \textit{et al.} \cite{unifyingHorne} giving a unified framework to describe these experiments. This further led to theoretical studies that examined the relationship between the one- and two-photon interference visibilities, through a beamsplitter \cite{bs_complementarity}, and through a double slit \cite{DS_complementarity} and established an inequality between one- and two-photon visibilities $V_m^2+V_\text{12}^2\le1$. Later, Saleh \textit{et al.} \cite{saleh_theory} expressed the equality case in terms of van Cittert–Zernike theorem \cite{goodman1985statistical} and showed the analogy between partial coherence and partial entanglement which they experimentally demonstrated for a double slit later \cite{saleh_exp}. Their experiment records the interference fringes using an ICCD at various distances between the double slit and the detector. Since then, several experiments have probed various properties of the spatially entangled photons. It was used by Neves \textit{et al.} to characterise the spatial entanglement present in the down-converted photons by measuring the Schmidt decomposition \cite{padua}. They also measure the concurrence (introduced by Wootters in \cite{concurrence}). Soon, building on this work, Peeters \textit{et al.} \cite{exter} showed a way to engineer the general 2-qubit state using a double slit by using imaging/Fourier optics after the double slit and by varying the slit separation. The effect of pump beam divergence on the fringes was qualitatively studied by Ryosuke \textit{et al.} without quantifying the visibilities or their complementarity \cite{waistNoVis}. The effect of entanglement on the diffraction envelope to the interference pattern from a double slit was also studied \cite{dsDiffraction}. The multi-mode nature of the SPDC was exploited in a double slit experiment that violated the which-slit and interference inequality \cite{whichPath, whichPath2}.

    In this paper, we demonstrate the duality between the one- and two-photon visibilities by continuously tweaking the number of independent emitters that participate in generating the entangled photons. We do so by varying the size of the pump waist falling on our nonlinear crystal and quantify the visibilities and its complementarity. We further report use cases for the various degrees of spatial entanglement that we generate and interpret our results using theoretical treatment that we develop in terms of the concept of the birth zone number introduced by Schneeloch \textit{et al.} \cite{birthzone}.

\section{Theory: The two-photon wavefunction}\label{sec:two-photon wavefunction}
    \subsection{The double Gaussian approximation}\label{subsec:dg}      
        When a down-converting nonlinear crystal is pumped by a Gaussian beam, the wavefunction of the collinearly down-converted SPDC photons at the crystal centre under certain approximations is given by \cite{law&eberly,birthzone}
        \begin{eqnarray}
            \Psi\left(\boldsymbol{q}_1, \boldsymbol{q}_2\right)=A \exp\left(-\frac{w_0^2\left|\boldsymbol{q}_1+\boldsymbol{q}_2\right|^2}{4}\right) \exp\left(-\frac{b^2\left|\boldsymbol{q}_1-\boldsymbol{q}_2\right|^2}{4}\right),\label{eq:DG_mom}\\
            \psi\left(\boldsymbol{x}_1, \boldsymbol{x}_2\right)=B \exp\left(-\frac{\left|\boldsymbol{x}_1+\boldsymbol{x}_2\right|^2}{4w_0^2}\right) \exp\left(-\frac{\left|\boldsymbol{x}_1-\boldsymbol{x}_2\right|^2}{4b^2}\right),\label{eq:DG_pos}
        \end{eqnarray}
        where $\boldsymbol{q}_i=(k_{ix},k_{iy})$ is the transverse-momentum and $\boldsymbol{x}_i=(x_i,y_i)$ is the transverse-position of the $i^\text{th}$ photon, $w_0$ is the pump beam waist, $b^2=L/3k_p$ is the crystal parameter with $L$ being the length of the crystal and $k_p$ the magnitude of the wavevector of the pump and $A$ and $B$ are the normalization constants. The form of the wavefunction in Eq.~\eqref{eq:DG_mom} is popularly known as the double Gaussian (DG) approximation and its many advantages have been well documented \cite{birthzone}. Equation \eqref{eq:DG_mom} represents the state in the transverse momentum basis and Eq.~\eqref{eq:DG_pos} in the transverse position basis. In both representations, the first exponential follows the form of the Gaussian pump while the second exponential comes from approximating the phase-matching function \cite{law&eberly}.
    
        By examining the form of $\Psi\left(\boldsymbol{q}_1, \boldsymbol{q}_2\right)$ we can see that there are two limiting cases, $w_0>>b$ and $w_0<<b$. In the first case, the first exponential forces the transverse momentum of the down-converted photons to be anti-paired whereas in the second one, the second exponential forces it to be paired. Since for a given crystal the parameter $b$ is fixed, the only way we can continuously tweak $\Psi\left(\boldsymbol{q}_1, \boldsymbol{q}_2\right)$ is via the beam waist. The first case is experimentally achieved by loosely focusing the pump. While one might manage to achieve the second case by tightly focusing the pump beam in thick crystals. However, if the goal is to engineer states with paired degrees of freedom, it is more convenient to work with the Fourier transform of the wavefunction $\psi\left(\boldsymbol{x}_1, \boldsymbol{x}_2\right)$. This works because the Fourier transform of a Gaussian with width $\sigma$ is a Gaussian with width $2/\sigma$. Thus, if the state is prepared with $w_0>>b$ then in the transverse spatial coordinates, the state  $\psi\left(\boldsymbol{x}_1, \boldsymbol{x}_2\right)$ will have the widths related as $1/w_0<<1/b$, leaving the wavefunction paired in the transverse position. Our experiment explores the properties of the wavefunction in the first case as we continuously tweak the beam waist $w_0$.

    \subsection{Structure of the entanglement: The birth zone number}\label{subsec:birthzone}
        One of the signatures of entanglement is that while the composite state residing in the original Hilbert space might be pure, the component states are mixed in at least two of its subspaces \cite{Landau1927}. This mixedness is essentially the measure of entanglement called entanglement entropy \cite{entanglementEntropy}. 
        A quantity called the Schmidt number $K$ is a measure of the dimensions of the entangled state \cite{law&eberly} which provides an estimate of the bandwidth of secure information that can be encoded in the state \cite{Cozzolino2019}. We have recently reported a rapid way to accurately measure it \cite{oemz}. 
        A closely related quantity to $K$ is the birth zone number $N$ 
        with $K=(N+1/N)^2/4$ \cite{birthzone}.

\begin{figure}[ht!]
            \centering\includegraphics[width=\linewidth,trim= 0cm 7cm 0cm 6.5cm,clip]{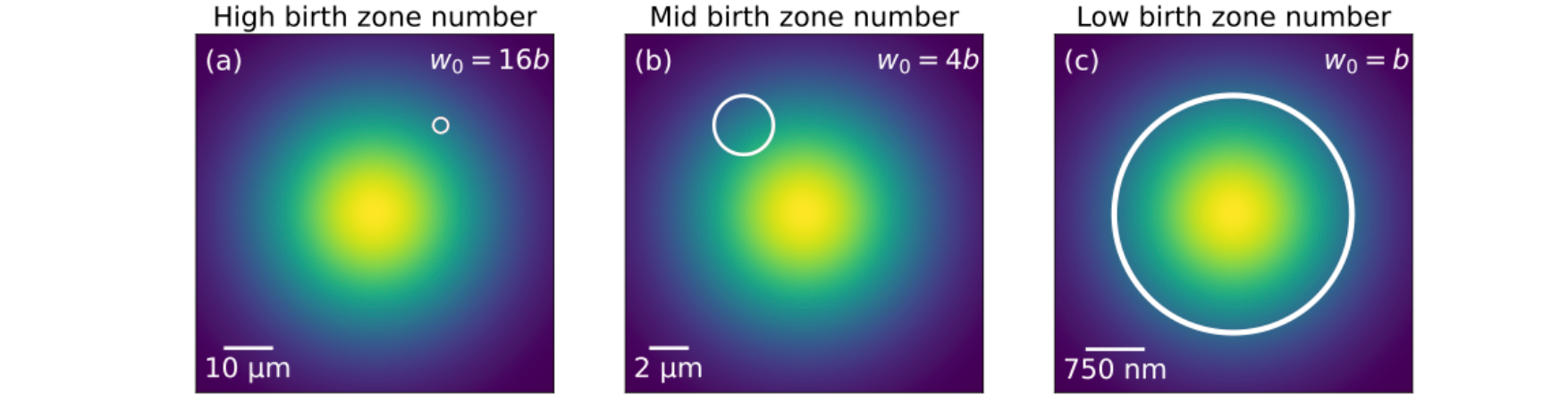}
            \caption{Simulations depicting a random birth zone for various birth zone numbers for (a) weakly focused light with $N=16$, (b) moderately focused light with $N=4$ and (c) tightly focused light with $N=1$. The white circle in each figure has a radius of $3~\mu$m. This shows how we can continuously tweak the number independently emitting centers by focusing the beam waist. }\label{fig:birthzone}
        \end{figure}

        Let us now assign a physical interpretation to the wavefunction $\psi\left(\boldsymbol{x}_1, \boldsymbol{x}_2\right)$ from Eq.~\eqref{eq:DG_pos}. The first part is dependent on $\left(\boldsymbol{x}_1+\boldsymbol{x}_2\right)/2$ which is the mean position of the two photons. This means that the centre of the two photons has an uncertainty of $w_0$ as it is proportional to the pump intensity. Similarly, the second part is dependent on $\left(\boldsymbol{x}_1-\boldsymbol{x}_2\right)/2$ which is the distance between the two photons. This means that the two photons are found roughly within a region of width $b$. This region is called the birth zone and since the two down-converted photons are, by definition, phase-matched, the length $b$ also is the coherence length of the photons emitted by the crystal. So, the emission region can be partitioned into blobs of area $\sim b^2$ that each emits independently. The birth zone number is defined as $N\equiv w_0/b$ and can be interpreted as the number of independent emitters targeted by the pump beam as seen in Fig.~\ref{fig:birthzone}.
    
        Since each birth zone emits independently, the photons they emit do not have a fixed phase relation. In fact, this feature of SPDC light was examined first in Ref.~\cite{SMmultiphotonMMsinglephoton} and later expanded in Ref.~\cite{coherenceEntanglementConnection} where the deep relation between partially entangled light and partially coherent light is established and used to measure the Schmidt number in the high entangled limit. Its angular spectrum was further studied in Ref.~\cite{angularcoherence}. It also enabled methods to study the entangled properties by measuring the coherence property of one of the pairs \cite{2from1_1,2from1_2,2from1_3,2from1_4}.

        We now draw attention to the fact that we can continuously tweak the number of emitters participating in the SPDC process by controlling the pump waist size falling on the crystal. When the pump is loosely focused, we are targeting a lot of emitters and thus, each photon in the down-converted light will be highly incoherent. On the other hand, when the pump waist is comparable to the birth zone size $b$, the down-converted light is coherent. 
        To see this, let us first express the wavefunction in terms of $N$ and $b$ 
        \begin{eqnarray}\label{eq:DG_birthzone}
            \Psi_N(\mathbf{q}_1,\mathbf{q}_2) = &A\text{exp}\left( -\frac{b^2(N^2+1)}{4} \left|\mathbf{q}_1\right|^2 \right)\text{exp}\left( -\frac{b^2(N^2+1)}{4}  \left|\mathbf{q}_2\right|^2 \right)\nonumber\\
            &\times\text{exp}\left( -\frac{b^2(N^2-1)}{2} \mathbf{q}_1\cdot\mathbf{q}_2 \right).
        \end{eqnarray}
        We can see that it is only the last exponential that entangles the two photons and it vanishes for $N=1$ making the two-photon wavefunction separable. For $N\to\infty$ we see that we can complete the square to get back a Gaussian in $\mathbf{q}_1+\mathbf{q}_2$.

    \section{Theory: Interference by a double slit}\label{subsec:ds}
        \subsection{The analytical approximation}\label{subsubsec:analytical}
\begin{figure}[ht]
                \centering
                \includegraphics[width=\linewidth,trim= 3.5cm 6.35cm 5.5cm 5.65cm,clip]{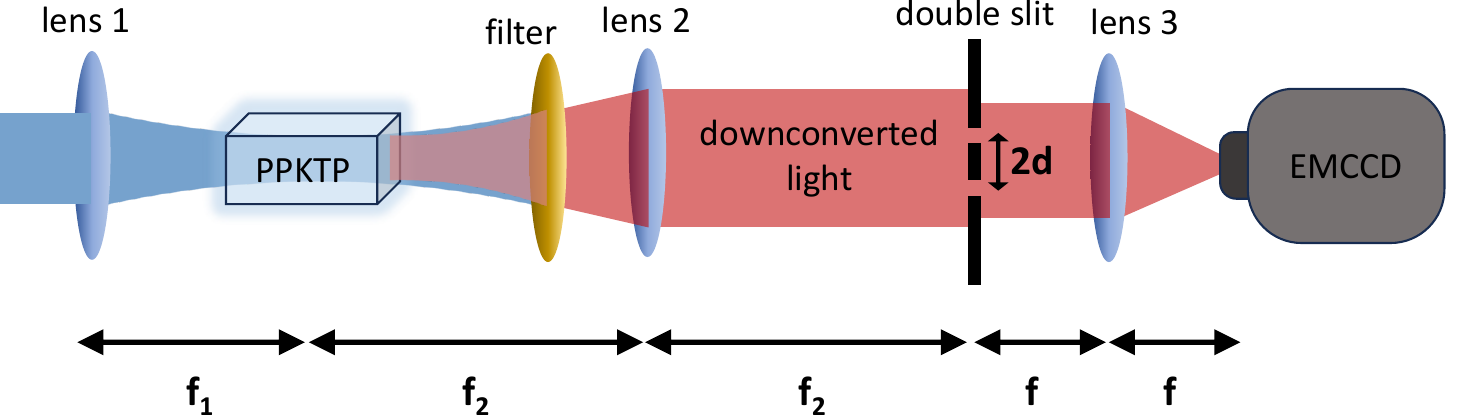}
                \caption{The Fourier plane of the down-converted light from a type-2 PPKTP is mapped onto the double slit using lens 2 (focal length $10$ cm). Its Fourier plane is then mapped onto the EMCCD by lens 3 (focal length $5$ cm).}
                \label{fig:setup}
            \end{figure}

            Since we established that the signature of entanglement is present in the coherence of the downconverted beam, to observe the effect of changing of pump width on entanglement, we resort to one of the simplest form of coherence measurement: the double slit. Figure \ref{fig:setup} shows the setup used in our experiment.

            To analytically treat the double slit interference, it is easier to consider slits 
            of infinitesimal openings that are placed symmetrically along the optic axis. In this case, the wavefunction can be written as \cite{exter}
             \begin{eqnarray}\label{eq:DS}
                |\psi_\text{DS}\rangle = \cos{\theta}\left( \frac{|L_1,R_2\rangle + |R_1,L_2\rangle}{\sqrt{2}} \right) + e^{i\varphi}\sin{\theta}\left( \frac{|L_1,L_2\rangle + |R_1,R_2\rangle}{\sqrt{2}} \right),
            \end{eqnarray}
             where the first(second) term corresponds to the case when the two photons are antipaired (paired) with $L_i~(R_i)$ corresponding to the $i^\text{th}$ photon passing through the left(right) slit. $\theta$ controls the degree to which the photons are paired based on Eq.~\eqref{eq:DG_mom} and $\varphi$ gives the phase difference between the two cases. By changing the pump waist, we can change $\theta$ and the location of the waist relative to the crystal centre decides the $\varphi$. Our experiment ran at a fixed $\varphi$ and $\pi/4\le\theta\le\pi/2$.

            The corresponding two-photon interference pattern is given by \cite{padua}
            \begin{eqnarray}\label{eq:ana_interference_pattern}
                G^{(2)}(x_1,x_2) \propto 1+V_{-}\cos(\phi_1-\phi_2)+V_{+}\cos(\phi_1+\phi_2)+V_m\left[ \cos(\phi_1) + \cos(\phi_2) \right],
            \end{eqnarray}
            where $V_-=\cos^2(\theta)$, $V_+=\sin^2(\theta)$ and $V_m=\sin(2\theta)\cos(\varphi)$ can be interpreted as visibilities of the fringes dependent on the variables $\phi_i=kdx_i/(2f)$ where $k$ is the magnitude of the wavevector of the down-converted photon, $2d$ is the slit separation, $f$ is the focal length of the lens after the double slit (see Fig.~\ref{fig:setup}). This tells us that in the coincidence space of $x_1$ vs. $x_2$, the diagonal and the anti-diagonal contain the two-photon information while the horizontal and vertical contain the marginal one-photon information. 

            We can readily see that $V_+ + V_- = 1$ which is a straightforward consequence of the normalisation of the state $\psi_\text{DS}$. Interestingly, Saleh \textit{et al.} in 2000 established a duality between the difference in the two two-photon visibilities ($V_\text{12}\equiv V_- - V_+$) and the marginal (for $\varphi=0$) as \cite{saleh_theory}
            \begin{eqnarray}\label{eq:12vis}
                V_\text{12}^2 + V_m^2 = 1.
            \end{eqnarray}
            In our experiment, the fringes from paired photons leading to $V_+$ come purely from one-photon contribution. Thus, $V_\text{12}$ can be interpreted as the visibility purely coming from two-photon correlation since the one-photon contribution along $V_-$ gets subtracted out by $V_+$. We interpret the duality in Eq.~\eqref{eq:12vis} as a consequence of the varying birth zone number. When $N=1$, the down-converted beam is coherent as the pump only activates one birth zone. This means that each photon of the entangled pair is independent of each other, and the resultant interference pattern should have high (low) visibility in real (coincidence) space. This corresponds to $\theta=\pi/4$ as it is equally likely that the photon is paired or anti-paired. When $\varphi=0$, the state can be factored into
            \begin{eqnarray}\label{eq:separable_state}
                |\psi_\text{DS,N=1}\rangle = \left(\frac{|L_1\rangle + |R_1\rangle}{\sqrt{2}}\right) \left(\frac{|L_2\rangle + |R_2\rangle}{\sqrt{2}}\right).
            \end{eqnarray}
            In this case, $G^\text{(2)}(x_1,x_2)$ will show a checkerboard pattern with $V_\text{12}=0 \text{ and } V_m=1$.
            
            On the other hand, when $N$ is large, the downconverted beam is highly incoherent as many birth zones are active. The interference fringes wash out in real space but will show bright fringes in the coincidence space as the wavefunction in the $x_1-x_2$ space now sees $N$ contributions. This corresponds to $\theta=0$ (although due to finite $b$, experimentally, this is never fully achieved) and the state is
            \begin{eqnarray}\label{eq:entangled_state}
                |\psi_{\text{DS,N}\to\infty}\rangle \to \frac{|L_1,R_2\rangle + |R_1,L_2\rangle}{\sqrt{2}}.
            \end{eqnarray}
            Here, $G^\text{(2)}(x_1,x_2)$ will show fringes along the diagonal with $V_\text{12}=1$ and $V_m=0$. For the intermediate cases, there will be fringes in the diagonal, the horizontal and the vertical directions. Although these relations between visibilities were derived for infinitesimal slit openings, it remains valid even for finite slit openings as this only causes an overall diffraction envelope leaving the interference fringe width and visibility untouched. But for a more accurate analysis, it is necessary to work out the wavefunction for a realistic double slit which is discussed in the following section.

        \subsection{The numerical estimation}\label{subsubsec:numerical}
            The wavefunction $\psi(x_1,x_2)$ of the down-converted photons at the crystal is given by Eq.~\eqref{eq:DG_pos}. We can see from our setup in Fig.~\ref{fig:setup} that lens 2 maps the momentum wavefunction $\Psi(q_1,q_2)$ onto the double slit with $q_i\equiv x_i\to 2\pi x_i/(f_2\lambda)$. The double slit aperture function for a slit separation of $2d$ and slit opening of $a$ is given by
            \begin{eqnarray}\label{eq:ds_function}
                \text{DS}(q;a,d) = \text{rect}\left(\frac{q+d}{a}\right) + \text{rect}\left(\frac{q-d}{a}\right),
            \end{eqnarray}
            where $\text{rect}$ is the rectangular function \cite{WOODWARD195326}. Thus the wavefunction just out of the double slit is given by
            \begin{eqnarray}\label{eq:ds_wavefunction}
                \Psi_\text{DS}(q_1,q_2) = \text{DS}(q_1;a,d)\text{DS}(q_2;a,d)~\Psi(q_1,q_2),
            \end{eqnarray}
            and the final lens maps this wavefunction out back into position space where the final wavefunction can be calculated using the convolution property of Fourier transform \cite{Arfken2013} as
            \begin{eqnarray}\label{eq:num_final_wavefunction}
                \psi(x_1,x_2) &=& \mathcal{FT}\left[\Psi_\text{DS}(q_1,q_2)\right], \nonumber \\
                             &=& \mathcal{FT} \left[\text{DS}(q_1;a,d)\text{DS}(q_2;a,d)\right]\star\mathcal{FT}\left[\Psi(q_1,q_2)\right],\\
                &=& 4a^2\cos\left(\frac{\pi d}{f\lambda}x_1\right)\text{sinc}\left(\frac{\pi a}{f\lambda}x_1\right) \cos\left(\frac{\pi d}{f\lambda}x_2\right) \text{sinc}\left(\frac{\pi a}{f\lambda}x+2\right) \star \psi\left(\frac{f}{f_2}x_1, \frac{f}{f_2}x_2\right),\nonumber
            \end{eqnarray}
            where $\mathcal{FT}[\cdot]$ represents the Fourier transform operation and $(\star)$ represents the convolution operation. The coupled nature of $x_1$ and $x_2$ in $\psi(x_1,x_2)$ makes it impossible to get a closed analytic form for the above convolution. Thus we numerically evaluate it whose results are discussed in the next section.

    \section{Experiment and Results}\label{sec:exp}
        \subsection{Experimental setup and method}
            The experimental setup as shown in Fig.~\ref{fig:setup} contains a $405$ nm Gaussian mode pump laser whose beam waist $w_0$ is made incident onto a $5$ mm thick type-2 periodically-poled Potassium Titanyl Phosphate (PPKTP) crystal by lens 1 with focal length $f_1$. We tweak $w_0$ by using various $f_1$s and measure it by a method described in \cite{oemz}. We image the propagating beam at various locations and fit a Gaussian propagation curve \cite{svelto_gaussian} to extract the waist. The temperature of the crystal is set to $40^\circ$C to ensure collinear phase matching. The excess pump is blocked by an interference filter. Next, the lens 2 ($f_2$ focal length) is kept at $f_2$ distance from the crystal centre which maps the momentum space wavefunction $\Psi_\text{DS}(q_1,q_2)$ onto the double slit which is kept in its back focal plane. The lens 3 of $f$ focal length ($5$ cm) is placed with the double slit ($2d=500\mu\text{m} \text{ and } a=150\mu\text{m}$) in its front focal plane. An EMCCD is placed in its back focal plane. We use an additional band-pass filter centred at $810$ nm of $3$ nm bandwidth. The EMCCD finally measures the interference pattern governed by $\left|\psi(x',x'')\right|^2$ from Eq.~\eqref{eq:num_final_wavefunction}. 

            We acquire 2M frames of photo-electron counts on the EMCCD with the settings: temperature of $-80^\circ$C, EM gain of 300, pre-amp gain of 3, horizontal readout rate of 5MHz, vertical shift speed of 0.5 $\mu$s and Vertical clock voltage amp of +1. To extract photon number from the photo-electron counts we use our multi-thresholding scheme described in Ref. \cite{Rounakpaper}. Next, with the extracted photon numbers we calculate the joint probability distribution (JPD) of photons between two pixels as
            \begin{eqnarray}\label{eq:jpd4d}
                \Gamma_{ijkl} = \langle c_{ij}c_{kl}\rangle - \langle c_{ij}\rangle\langle c_{kl}\rangle,
            \end{eqnarray}
            where $c_{ij}$ is the photon counts falling on the pixel centered at $(x_i,y_j)$ and $\langle\cdot\rangle$ indicates an ensemble average over the total frames. Each pixel is $16\mu\text{m}\times16\mu\text{m}$ in size. The second term in Eq.~\eqref{eq:jpd4d} statistically accounts for the accidental coincidences.
        
            In our coordinate system, $x$ is the direction perpendicular to the slits, $y$ is along the slits and $z$ is the propagation direction. Since our EMCCD captures 2D images, the JPD is a 4D object. Since we are only interested in the direction where the double slits cause fringing so we average over $y$ to get
            \begin{eqnarray}\label{eq:jpd2d}
                \left|\psi(x_i,x_k)\right|^2 = \Gamma_{ik} \equiv \sum_{j=1}^{N_y}\sum_{l=1}^{N_y} \Gamma_{ijkl}
            \end{eqnarray}
            where $N_x$ ($N_y$)is the number of pixels of the frames along $x$ ($y$) direction. Since we want to observe the fringes in $x_1$ (equivalently $x_2$) and $x_1-x_2$, it is useful to calculate the marginal and the correlation function. These are calculated from the JPD as follows
            \begin{eqnarray}
                \rho(x) &=& \sum_{j=1}^{N_y}\Gamma_{ijij},\label{eq:marginal}\\
                \text{Corr}(X) &=& \sum_{j=1}^{N_y}\sum_{l=1}^{N_y}\left(\sum_{i=1}^{N_x}\Gamma_{ij(i-X)l}\right)\label{eq:correlation}.
            \end{eqnarray}

        \subsection{Results}
\begin{figure}[ht]
            \centering    
            \includegraphics[width=\linewidth]{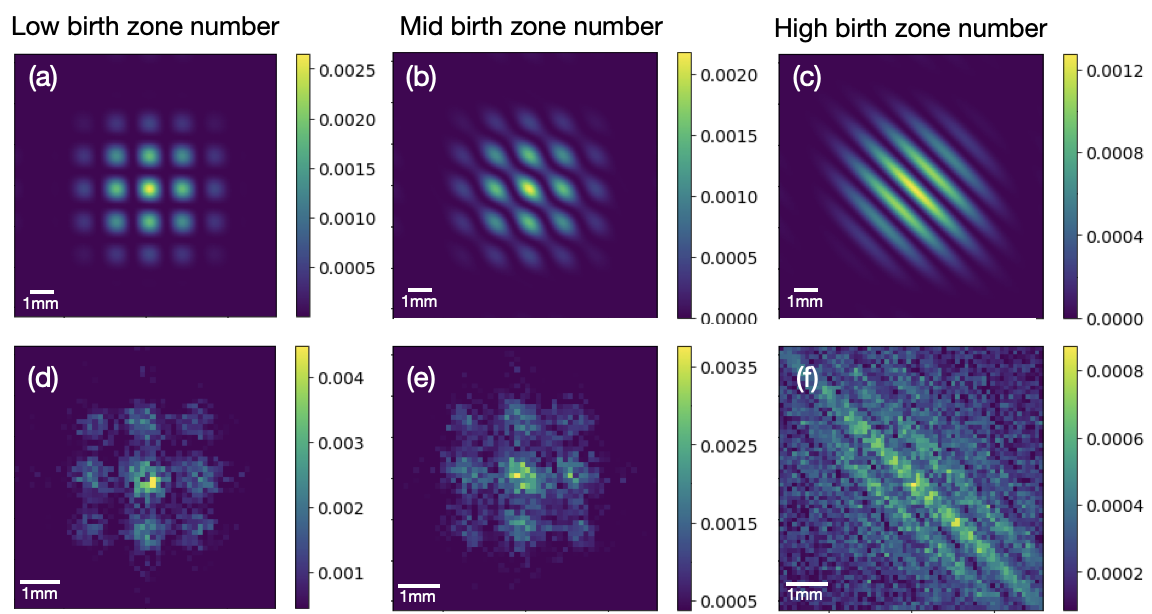}
            \caption{ Images (a)-(c) show the calculated 2D JPD (using Eq.~\eqref{eq:jpd2d}) for three regimes of the birth zone number. The images (d)-(f) show the numerically calculated 2D JPD from Eq.~\eqref{eq:num_final_wavefunction} as the wavefunction with $w_0$ set correspondingly to the experimentally evaluated values.}
            \label{fig:jpd}
        \end{figure}

            Since all our calculations are based on the JPD, we are in the subspace when both of the entangled photons make it through to the detector. This situation is indifferent to the contribution coming from mixed states with partially lost photons. The numerically estimated JPDs using Eq.~\eqref{eq:num_final_wavefunction} for $N\approx10,17,34$ are shown in Fig.~\ref{fig:jpd}a, b and c respectively. The experimentally extracted JPDs for the same $N$s using  Eq.~\eqref{eq:jpd2d} are shown in Fig.~\ref{fig:jpd}d, e and f for comparison. 

            We can see that for low $N$ the fringes are purely in the horizontal and vertical directions indicating a high $V_m$ and low $V_\text{12}$. As discussed in Sec.~\ref{subsec:birthzone}, this indicates the high coherence present in the down-converted beam. For experiments where SPDC is used as a heralded single photon source or for its polarization entanglement, it is usually detected in fibre-coupled detectors. In these scenarios where the spatial entanglement is not used, it is advantageous to work at low $N$ as their coherence ensures that their mode can be matched easily to that of the fibre using lenses. 

            In the mid $N$ plot (Fig.~\ref{fig:jpd}b \& e) we see that the fringes along the diagonal start to appear which reduces the visibility along the horizontal and vertical. These states show some properties of spatial entanglement while still retaining a good coupling with fibres. These states are ideal for the unique detection scheme demonstrated by Leach \textit{et al.} in Ref. \cite{space2timeDetector} where they use a spatial array of fibres of different lengths that map spatial location to time that all fall into a single detector. For high $N$ (Fig.~\ref{fig:jpd}c \& f) we see absolutely no fringes in the horizontal and vertical direction but we see clear fringes along the diagonal indicating that these states are highly spatially entangled.

        \begin{figure}
            \centering    
            \includegraphics[width=\linewidth,trim= 0.1cm 0cm 0cm 0cm,clip]{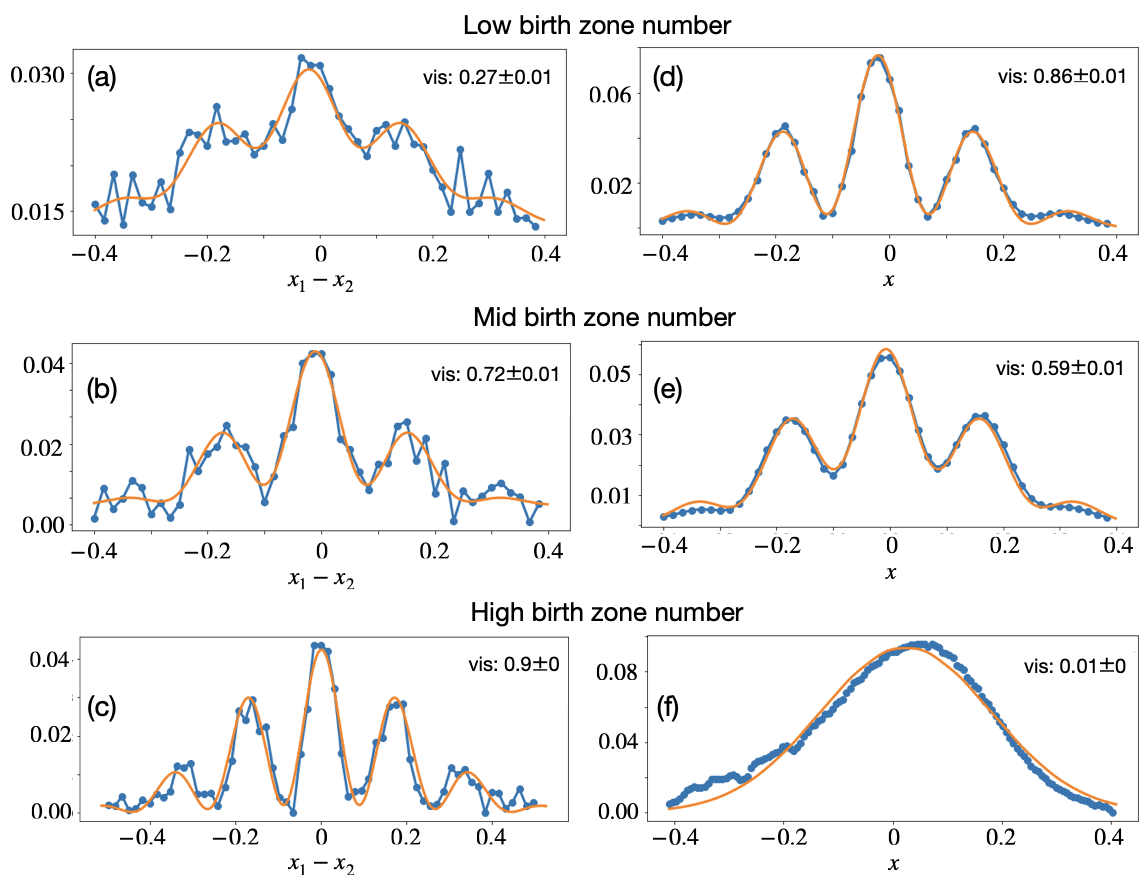}
            \caption{Fringes are calculated from correlation for three regimes of the birth zone number. The plots (a)-(c) show the measured $\text{Corr}(x_1-x_2)$ where the (two-photon) fringes are present in correlation space while the plots (d)-(f) show the measured $\rho(x)$ where the (one-photon) are fringes in real space. The blue points are experimentally measured (using Eqs.~\eqref{eq:correlation} and \eqref{eq:marginal}) and the orange curves are their fits with Eq.~\eqref{eq:fitfunc} as the function. The visibilites calculated from the fit are given on the corresponding plots with their fit covariances.}
            \label{fig:fringes}
        \end{figure}

            The one- and two-photon fringes as calculated from the JPD using Eqs.~\eqref{eq:marginal} and \eqref{eq:correlation} respectively are shown for the same three birth zone numbers $N\approx10,17,34$ as blue dots in Fig.~\ref{fig:fringes}. To fit this data we use a fit function of the form \cite{padua}
            \begin{eqnarray}\label{eq:fitfunc}
                f(t) = A\text{sinc}^2\left( \frac{\pi a}{f\lambda}t\right)\left[ 1 + V\cos\left(\frac{2\pi d}{f\lambda} t\right) \right] + c,
            \end{eqnarray}
            where $t=x_1=x_2$ and $V=V_m$ for fitting the marginal data and $t=x_1-x_2$ and $V=V_\text{12}$ for fitting the correlation data with the amplitude $A$, the offset $c$ and the visibility $V$ as the free parameters. The $\text{sinc}$ part comes from the diffraction through the finite slit openings $a$ while the $\cos$ part comes from the interference between the two slits. The correlation data is noisy because the correction for accidental coincidences works only statistically as it only subtracts the mean contribution of the beam.

\begin{figure}
            \centering
            \includegraphics[width=\linewidth,trim= 1.75cm 0cm 3cm 0cm,clip]{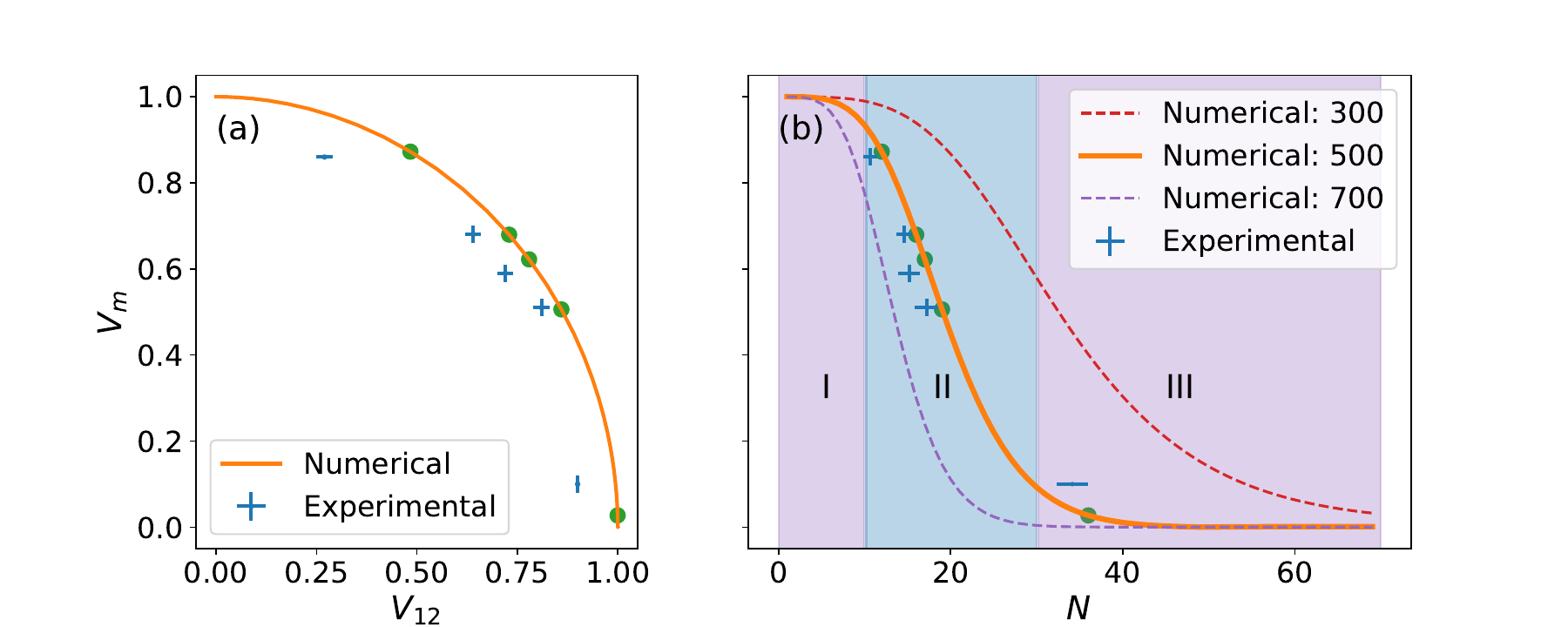}
            \caption{The complementarity between the one- and two-photon visibilities is shown in (a) where the experimental points (blue with error bars in $V_m$ and $V_\text{12}$) follow the circular trend as predicted by Eq.~\eqref{eq:12vis}). The orange curve is formed by numerical evaluations of the visibilities with the green dots marking the positions corresponding to the experimental parameters. The deviation is due to the sensitivity of $V_\text{12}$ on the exact location of the beam waist on the crystal. (b) shows $V_m$ as a function of the birth zone number depicting the loss of coherence. The colour coding remains the same as in (a). The red (purple) dashed curves show the numerically estimated $V_m$ for slit separation of $300\mu$m ($700\mu$m). This shows that there is a region of rapid fall in the one-photon visibility and on either side of that it is flat. The three regimes of entanglement are depicted in the shaded regions I, II and III.}
            \label{fig:visibility}
        \end{figure}

            Figure~\ref{fig:visibility}a demonstrates the complementarity between the one- and two-photon interference measured experimentally using visibilities extracted from Eq.~\eqref{eq:fitfunc}. We suspect the discrepancy between the measured visibilities from the theoretical prediction of unity is due to the sensitivity of the setup to the location of the beam waist with respect to the crystal center. The non-zero $\varphi$ (non-flat phase-front) at the crystal centre transfers some entanglement from space to phase part of our state  \cite{padua,exter,qualityOfPhase}. Figure~\ref{fig:visibility}b shows, that for the first few $N$s, the one-photon visibility doesn't change appreciably. There after it rapidly falls before being flat again. This indicates that the loss of spatial coherence (correspondingly the gain of spatial entanglement) saturates after a certain size of the beam waist. This depends on the length-scale at which we are probing the spatial entanglement which, in our case, is set by the slit separation $2d$. A large $d$ ensures that the photons are mostly anti-paired at the slits at the cost of low fringe width while a smaller $d$ has larger fringe width while being contaminated with paired contributions. This can be seen in Fig.~\ref{fig:visibility}b from our numerical evaluation for fringes with $d=300 \text{ and } 700\mu$m. Such a behaviour is strongly relevant to ghost imaging and is discussed in more detail in Ref.~\cite{2regionsOfgrowth}. 

    \section{Conclusion}\label{sec:conclusion}

        In conclusion, we studied the duality between coherence and entanglement of down-converted wavefunction described by a double Gaussian (Eq.~\eqref{eq:DG_mom}). We infer these properties from the one- and two-photon visibilities of interference fringes due to a double slit for various beam waists at the crystal centre. We show how we extract information from the calculated JPD (Eq.~\eqref{eq:jpd4d}) from the multi-thresholded images inferred from our EMCCD. From these, we examine and demonstrate the duality between the one- and two-photon interference visibilities evaluated using Eq.~\eqref{eq:fitfunc}. We interpret the transfer of coherence from the individual photons to coherence between the entangled photons as a consequence of the changing birth zone number. 
        
Examining the entanglement in terms of the birth zone number allows us to identify three regimes of entanglement, namely, low, mid and high, each one having its optimal place in quantum experiments. We argue how the low regime is ideal for fiber-coupled photon counting experiments. The mid regime is ideal for studying spatial entanglement that employ scanning fiber-coupled single-photon detectors. The high regime is best for studying spatial entanglement using cameras such as EMCCDs. We believe that our quick and easy method of varying the pump waist can help scan between these distinct regimes to find the optimal one for each experimental scenario. This observation and ability will have a noticeable impact on the wide variety of quantum optical experiments globally.
        
    \begin{backmatter}
        \bmsection{Funding}
            Department of Atomic Energy, Government of India Project Identification No. RTI4002 under DAE OM (1303/1/2020/R\&D-II/DAE/5567).
    
        \bmsection{Disclosures}
            The authors declare no conflicts of interest.
    
        \bmsection{Data availability} 
            Data underlying the results presented in this paper are not publicly available at this time but may be obtained from the authors upon reasonable request.
    \end{backmatter}

    \bibliography{sample}

\end{document}